\documentclass[twocolumn]{article}

\title{Ants are not Conscious}
\author{Russell K. Standish\\
School of Mathematics and Statistics\\
The University of New South Wales, Australia\\
email: hpcoder@hpcoders.com.au
}

\usepackage{amssymb, natbib}
\newcommand{\naturals}{{\mathbb N}}

\usepackage{pstricks,pslatex}

\begin{document}

\maketitle

\begin{abstract}
Anthropic reasoning is a form of statistical reasoning based upon
finding oneself a member of a particular reference class of conscious
beings. By considering empirical distribution functions defined over
animal life on Earth, we can deduce that the vast bulk of animal life
is unlikely to be conscious. 

\end{abstract}

\section*{Introduction}

Consciousness is a {\em b\^ete noir} of the physical sciences. Each
and everyone of us is aware of his or her own consciousness, and
indeed it seems to be necessary in order to carry out science, or at
very least to give meaning to its theories and results. Yet, the more
neurophysiologists probe the workings of the brain, the more of a
phantom consciousness appears to be. Some argue that even if a
complete neurophysical theory of the brain's function is determined,
the ``hard'' problem of how phenomenal experience is generated still
remains \citep{Chalmers95}. Related to this issue is that we cannot prove
definitively that any other individual of the human race is conscious
and not a zombie that acts for all intents and purposes as
conscious. Consciousness is fundamentally a first-person phenomenon
with scientific discourse relegated to comparing reports with our own
experience. Yet it is unreasonable to doubt the consciousness of other
humans, who are constructed in the same way as ourselves, and who act
in the same way as ourselves. The same is not true of other species,
who are constructed from different body plans, have very different
neural structures, and act in significantly different ways to
ourselves. In the words of \citet{Nagel74} ``What is it like to be a
bat?'', answering the question of consciousness in animals
seems hopeless. Whilst Nagel was assuming that it is something to be
like a bat, it is entirely reasonable to ask the question of whether it
is anything to be like a bat. Most people would assume that on a scale
of organism complexity from human beings, through vertebrates,
invertebrates, etc. through to non-living matter, a line can be drawn
between organisms experiencing phenomenal consciousness and those that
don't. Descartes, for example, drew the line between humans and
non-humans. Others would argue that some other species of mammal, and
possibly bird as well as some cephalopods are probably conscious. Some
even argue that insects might be conscious \citep{Tye97}. 

In this paper, I define consciousness in an operational way by noting
that the {\em reference class} of anthropic reasoning \citep{Bostrom02}
must consist of conscious entities, possibly restricted in some way,
such as the set of terrestrial animals. 
Anthropic reasoning is best
known in the form of the Cosmological Anthropic
Principle \citep{Barrow-Tipler86} and the infamous Doomsday
Argument \citep{Leslie89}.

Anthropic reasoning has been criticised on a number of fronts,
particularly where it has been applied to produce counter intuitive
conclusions. For example, the fine tuning argument has been used as
evidence for a divine creator, or as evidence for a Multiverse, and
the doomsday argument suggests that the human population will crash in
the not too distant future. Most of these objections have been
rebutted in Bostrom's book \citep{Bostrom02}, who makes a well-argued case
that anthropic reasoning can be done validly. It is not the purpose of
this paper to review to the structure of these arguments, objections
raised, nor rebuttals of those objections, as that is incidental to
the aims of this paper. However, two issues in particular are
pertinent: the reference class problem, and the measure problem.

The issue of what constitutes
the class of observers from which the subject observer reasons he/she
was randomly sampled is known as the reference class problem. For many
examples of Anthropic Reasoning, precisely what constitutes the
reference class does not bear much on the conclusions of the
reasoning. \citet{Bostrom02} gives examples of this. In this
paper, we very much turn the reasoning on it head, and ask what can we
establish about the reference class, given the observation of what we
are, and other information we might have at hand. 

It might be argued that the reference class used for anthropic
reasoning should only include those observers capable of understanding
anthropic arguments, or more widely, those conscious entities capable
of introspection. It is not at all clear whether this would include
all humans, just a subset of humans, or non-human species as
well. Conversely, the widest possible reference class is the set of
all conscious observers, the interpretation I wish to use here. An
alternative reading of this paper is that it is not talking about
consciousness per se, but what is, or is not, allowable within the
anthropic reference class.

The measure problem comes from extending anthropic reasoning to
infinite sets of observers, such as we would expect to be the case in
a Multiverse. In the set $\naturals=\{0,1,2,\ldots\}$, we might be
tempted to say that the set of even numbers has measure 0.5. Yet if we
write the set in a different order as
$\naturals=\{0,1,3,5,2,7,9,11,4,\ldots\}$, the same line of argument
produces a measure of 0.25. However, with respect to the arguments
given in this paper, this measure problem doesn't arise, as the
measure is already known empirically.

To consider the title question of this paper, we only need to note
that there are considerably more ants than humans in the world. It is
estimated that ants monopolise between 15--20\% of the 
terrestrial animal biomass \citep{Schultz00}, far exceeding that of the
vertebrates. A typical suburban house garden will contain city-scale
populations of ants. A na\"\i{}ve application of anthropic reasoning
would conclude that ants could not be conscious, as otherwise we would
expect to be an ant rather than a human.

Unfortunately this usage of anthropic reasoning raises the Chinese
paradox. Why wasn't I born in the most populous nation on Earth,
China, which has 50 times the population of my country of birth,
Australia? 

Furthermore, ants are not a single species, but are taxonomically
speaking a family, with which we are comparing a single species {\em
  homo sapiens}. It would be better to rephrase the question in a way
that didn't depend on a somewhat human-biased taxonomic scheme.

In the rest of this paper I show that the Chinese paradox is actually
not a problem for anthropic reasoning, and in so doing demonstrate a
previously unknown process for generating power law distributions.
Then by recasting the ant consciousness problem into a question of
expected body mass, which is an objective physical measurement rather
than a possibly subjective classification, and including the proper
handling of the Chinese paradox, we can conclude that the vast
majority of animal species (particularly the small ones) are unlikely
to be conscious (or in the reference class, if you prefer) by
anthropic reasoning.

\section*{Chinese paradox}\label{chinese}

China has
well over a billion people, and along with India, has by far the
biggest population of all the nations in the world. I happen
to live in Australia, for instance, a country with around 1/50th the
population of China. It would be absurd to conclude that
Chinese\index{Chinese argument}
people are unconscious, so na\"\i{}vely one would expect on anthropic grounds to be
Chinese or Indian. At first sight this looks disastrous for anthropic
reasoning, until you realise that it is ill-posed. Suppose you asked
the question of what is the chance of being Chinese versus not being
Chinese. There is about a 20\% chance of being Chinese, and 80\% not,
so it then becomes unsurprising to not be Chinese.

\begin{figure*}
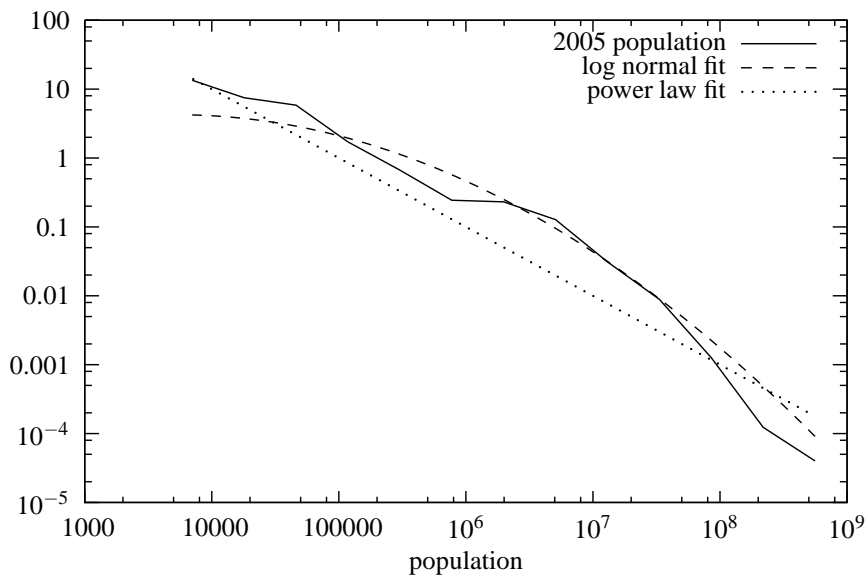

\begin{center}
\ifx\PSTloaded\undefined
\def\PSTloaded{t}
\psset{arrowsize=.01 3.2 1.4 .3}
\psset{dotsize=.01}
\catcode`@=11

\newpsobject{PST@Border}{psline}{linewidth=.0015,linestyle=solid}
\newpsobject{PST@Axes}{psline}{linewidth=.0015,linestyle=dotted,dotsep=.004}
\newpsobject{PST@Solid}{psline}{linewidth=.0015,linestyle=solid}
\newpsobject{PST@Dashed}{psline}{linewidth=.0015,linestyle=dashed,dash=.01 .01}
\newpsobject{PST@Dotted}{psline}{linewidth=.0025,linestyle=dotted,dotsep=.008}
\newpsobject{PST@LongDash}{psline}{linewidth=.0015,linestyle=dashed,dash=.02 .01}
\newpsobject{PST@Diamond}{psdots}{linewidth=.001,linestyle=solid,dotstyle=square,dotangle=45}
\newpsobject{PST@Filldiamond}{psdots}{linewidth=.001,linestyle=solid,dotstyle=square*,dotangle=45}
\newpsobject{PST@Cross}{psdots}{linewidth=.001,linestyle=solid,dotstyle=+,dotangle=45}
\newpsobject{PST@Plus}{psdots}{linewidth=.001,linestyle=solid,dotstyle=+}
\newpsobject{PST@Square}{psdots}{linewidth=.001,linestyle=solid,dotstyle=square}
\newpsobject{PST@Circle}{psdots}{linewidth=.001,linestyle=solid,dotstyle=o}
\newpsobject{PST@Triangle}{psdots}{linewidth=.001,linestyle=solid,dotstyle=triangle}
\newpsobject{PST@Pentagon}{psdots}{linewidth=.001,linestyle=solid,dotstyle=pentagon}
\newpsobject{PST@Fillsquare}{psdots}{linewidth=.001,linestyle=solid,dotstyle=square*}
\newpsobject{PST@Fillcircle}{psdots}{linewidth=.001,linestyle=solid,dotstyle=*}
\newpsobject{PST@Filltriangle}{psdots}{linewidth=.001,linestyle=solid,dotstyle=triangle*}
\newpsobject{PST@Fillpentagon}{psdots}{linewidth=.001,linestyle=solid,dotstyle=pentagon*}
\newpsobject{PST@Arrow}{psline}{linewidth=.001,linestyle=solid}
\catcode`@=12

\fi
\psset{unit=5.0in,xunit=5.0in,yunit=3.0in}
\pspicture(0.000000,0.000000)(1.000000,1.000000)
\ifx\nofigs\undefined
\catcode`@=11

\PST@Border(0.1490,0.1260)
(0.1640,0.1260)

\PST@Border(0.9470,0.1260)
(0.9320,0.1260)

\rput[r](0.1330,0.1260){$10^{-5}$}
\PST@Border(0.1490,0.1622)
(0.1565,0.1622)

\PST@Border(0.9470,0.1622)
(0.9395,0.1622)

\PST@Border(0.1490,0.2101)
(0.1565,0.2101)

\PST@Border(0.9470,0.2101)
(0.9395,0.2101)

\PST@Border(0.1490,0.2346)
(0.1565,0.2346)

\PST@Border(0.9470,0.2346)
(0.9395,0.2346)

\PST@Border(0.1490,0.2463)
(0.1640,0.2463)

\PST@Border(0.9470,0.2463)
(0.9320,0.2463)

\rput[r](0.1330,0.2463){$10^{-4}$}
\PST@Border(0.1490,0.2825)
(0.1565,0.2825)

\PST@Border(0.9470,0.2825)
(0.9395,0.2825)

\PST@Border(0.1490,0.3304)
(0.1565,0.3304)

\PST@Border(0.9470,0.3304)
(0.9395,0.3304)

\PST@Border(0.1490,0.3549)
(0.1565,0.3549)

\PST@Border(0.9470,0.3549)
(0.9395,0.3549)

\PST@Border(0.1490,0.3666)
(0.1640,0.3666)

\PST@Border(0.9470,0.3666)
(0.9320,0.3666)

\rput[r](0.1330,0.3666){ 0.001}
\PST@Border(0.1490,0.4028)
(0.1565,0.4028)

\PST@Border(0.9470,0.4028)
(0.9395,0.4028)

\PST@Border(0.1490,0.4506)
(0.1565,0.4506)

\PST@Border(0.9470,0.4506)
(0.9395,0.4506)

\PST@Border(0.1490,0.4752)
(0.1565,0.4752)

\PST@Border(0.9470,0.4752)
(0.9395,0.4752)

\PST@Border(0.1490,0.4869)
(0.1640,0.4869)

\PST@Border(0.9470,0.4869)
(0.9320,0.4869)

\rput[r](0.1330,0.4869){ 0.01}
\PST@Border(0.1490,0.5231)
(0.1565,0.5231)

\PST@Border(0.9470,0.5231)
(0.9395,0.5231)

\PST@Border(0.1490,0.5709)
(0.1565,0.5709)

\PST@Border(0.9470,0.5709)
(0.9395,0.5709)

\PST@Border(0.1490,0.5955)
(0.1565,0.5955)

\PST@Border(0.9470,0.5955)
(0.9395,0.5955)

\PST@Border(0.1490,0.6071)
(0.1640,0.6071)

\PST@Border(0.9470,0.6071)
(0.9320,0.6071)

\rput[r](0.1330,0.6071){ 0.1}
\PST@Border(0.1490,0.6434)
(0.1565,0.6434)

\PST@Border(0.9470,0.6434)
(0.9395,0.6434)

\PST@Border(0.1490,0.6912)
(0.1565,0.6912)

\PST@Border(0.9470,0.6912)
(0.9395,0.6912)

\PST@Border(0.1490,0.7158)
(0.1565,0.7158)

\PST@Border(0.9470,0.7158)
(0.9395,0.7158)

\PST@Border(0.1490,0.7274)
(0.1640,0.7274)

\PST@Border(0.9470,0.7274)
(0.9320,0.7274)

\rput[r](0.1330,0.7274){ 1}
\PST@Border(0.1490,0.7636)
(0.1565,0.7636)

\PST@Border(0.9470,0.7636)
(0.9395,0.7636)

\PST@Border(0.1490,0.8115)
(0.1565,0.8115)

\PST@Border(0.9470,0.8115)
(0.9395,0.8115)

\PST@Border(0.1490,0.8361)
(0.1565,0.8361)

\PST@Border(0.9470,0.8361)
(0.9395,0.8361)

\PST@Border(0.1490,0.8477)
(0.1640,0.8477)

\PST@Border(0.9470,0.8477)
(0.9320,0.8477)

\rput[r](0.1330,0.8477){ 10}
\PST@Border(0.1490,0.8839)
(0.1565,0.8839)

\PST@Border(0.9470,0.8839)
(0.9395,0.8839)

\PST@Border(0.1490,0.9318)
(0.1565,0.9318)

\PST@Border(0.9470,0.9318)
(0.9395,0.9318)

\PST@Border(0.1490,0.9563)
(0.1565,0.9563)

\PST@Border(0.9470,0.9563)
(0.9395,0.9563)

\PST@Border(0.1490,0.9680)
(0.1640,0.9680)

\PST@Border(0.9470,0.9680)
(0.9320,0.9680)

\rput[r](0.1330,0.9680){ 100}
\PST@Border(0.1490,0.1260)
(0.1490,0.1460)

\PST@Border(0.1490,0.9680)
(0.1490,0.9480)

\rput(0.1490,0.0840){ 1000}
\PST@Border(0.1890,0.1260)
(0.1890,0.1360)

\PST@Border(0.1890,0.9680)
(0.1890,0.9580)

\PST@Border(0.2420,0.1260)
(0.2420,0.1360)

\PST@Border(0.2420,0.9680)
(0.2420,0.9580)

\PST@Border(0.2691,0.1260)
(0.2691,0.1360)

\PST@Border(0.2691,0.9680)
(0.2691,0.9580)

\PST@Border(0.2820,0.1260)
(0.2820,0.1460)

\PST@Border(0.2820,0.9680)
(0.2820,0.9480)

\rput(0.2820,0.0840){ 10000}
\PST@Border(0.3220,0.1260)
(0.3220,0.1360)

\PST@Border(0.3220,0.9680)
(0.3220,0.9580)

\PST@Border(0.3750,0.1260)
(0.3750,0.1360)

\PST@Border(0.3750,0.9680)
(0.3750,0.9580)

\PST@Border(0.4021,0.1260)
(0.4021,0.1360)

\PST@Border(0.4021,0.9680)
(0.4021,0.9580)

\PST@Border(0.4150,0.1260)
(0.4150,0.1460)

\PST@Border(0.4150,0.9680)
(0.4150,0.9480)

\rput(0.4150,0.0840){ 100000}
\PST@Border(0.4550,0.1260)
(0.4550,0.1360)

\PST@Border(0.4550,0.9680)
(0.4550,0.9580)

\PST@Border(0.5080,0.1260)
(0.5080,0.1360)

\PST@Border(0.5080,0.9680)
(0.5080,0.9580)

\PST@Border(0.5351,0.1260)
(0.5351,0.1360)

\PST@Border(0.5351,0.9680)
(0.5351,0.9580)

\PST@Border(0.5480,0.1260)
(0.5480,0.1460)

\PST@Border(0.5480,0.9680)
(0.5480,0.9480)

\rput(0.5480,0.0840){$10^{6}$}
\PST@Border(0.5880,0.1260)
(0.5880,0.1360)

\PST@Border(0.5880,0.9680)
(0.5880,0.9580)

\PST@Border(0.6410,0.1260)
(0.6410,0.1360)

\PST@Border(0.6410,0.9680)
(0.6410,0.9580)

\PST@Border(0.6681,0.1260)
(0.6681,0.1360)

\PST@Border(0.6681,0.9680)
(0.6681,0.9580)

\PST@Border(0.6810,0.1260)
(0.6810,0.1460)

\PST@Border(0.6810,0.9680)
(0.6810,0.9480)

\rput(0.6810,0.0840){$10^{7}$}
\PST@Border(0.7210,0.1260)
(0.7210,0.1360)

\PST@Border(0.7210,0.9680)
(0.7210,0.9580)

\PST@Border(0.7740,0.1260)
(0.7740,0.1360)

\PST@Border(0.7740,0.9680)
(0.7740,0.9580)

\PST@Border(0.8011,0.1260)
(0.8011,0.1360)

\PST@Border(0.8011,0.9680)
(0.8011,0.9580)

\PST@Border(0.8140,0.1260)
(0.8140,0.1460)

\PST@Border(0.8140,0.9680)
(0.8140,0.9480)

\rput(0.8140,0.0840){$10^{8}$}
\PST@Border(0.8540,0.1260)
(0.8540,0.1360)

\PST@Border(0.8540,0.9680)
(0.8540,0.9580)

\PST@Border(0.9070,0.1260)
(0.9070,0.1360)

\PST@Border(0.9070,0.9680)
(0.9070,0.9580)

\PST@Border(0.9341,0.1260)
(0.9341,0.1360)

\PST@Border(0.9341,0.9680)
(0.9341,0.9580)

\PST@Border(0.9470,0.1260)
(0.9470,0.1460)

\PST@Border(0.9470,0.9680)
(0.9470,0.9480)

\rput(0.9470,0.0840){$10^{9}$}
\PST@Border(0.1490,0.9680)
(0.1490,0.1260)
(0.9470,0.1260)
(0.9470,0.9680)
(0.1490,0.9680)

\rput(0.5480,0.0210){population}
\rput[r](0.8200,0.9270){2005 population}
\PST@Solid(0.8360,0.9270)
(0.9150,0.9270)

\PST@Solid(0.2615,0.8632)
(0.2615,0.8632)
(0.3158,0.8327)
(0.3702,0.8197)
(0.4245,0.7556)
(0.4788,0.7064)
(0.5332,0.6537)
(0.5875,0.6509)
(0.6419,0.6198)
(0.6962,0.5461)
(0.7505,0.4807)
(0.8049,0.3787)
(0.8592,0.2572)
(0.9135,0.1985)

\rput[r](0.8200,0.8850){log normal fit}
\PST@Dashed(0.8360,0.8850)
(0.9150,0.8850)

\PST@Dashed(0.2615,0.8026)
(0.2615,0.8026)
(0.2681,0.8023)
(0.2747,0.8018)
(0.2813,0.8013)
(0.2878,0.8006)
(0.2944,0.7998)
(0.3010,0.7989)
(0.3076,0.7979)
(0.3142,0.7968)
(0.3208,0.7956)
(0.3274,0.7943)
(0.3339,0.7929)
(0.3405,0.7913)
(0.3471,0.7897)
(0.3537,0.7879)
(0.3603,0.7861)
(0.3669,0.7841)
(0.3735,0.7820)
(0.3800,0.7798)
(0.3866,0.7775)
(0.3932,0.7751)
(0.3998,0.7726)
(0.4064,0.7699)
(0.4130,0.7672)
(0.4196,0.7644)
(0.4262,0.7614)
(0.4327,0.7584)
(0.4393,0.7552)
(0.4459,0.7519)
(0.4525,0.7485)
(0.4591,0.7450)
(0.4657,0.7414)
(0.4723,0.7377)
(0.4788,0.7339)
(0.4854,0.7299)
(0.4920,0.7259)
(0.4986,0.7218)
(0.5052,0.7175)
(0.5118,0.7131)
(0.5184,0.7087)
(0.5249,0.7041)
(0.5315,0.6994)
(0.5381,0.6946)
(0.5447,0.6897)
(0.5513,0.6846)
(0.5579,0.6795)
(0.5645,0.6743)
(0.5710,0.6689)
(0.5776,0.6635)
(0.5842,0.6579)
(0.5908,0.6523)
(0.5974,0.6465)
(0.6040,0.6406)
(0.6106,0.6346)
(0.6172,0.6285)
(0.6237,0.6223)
(0.6303,0.6160)
(0.6369,0.6095)
(0.6435,0.6030)
(0.6501,0.5963)
(0.6567,0.5896)
(0.6633,0.5827)
(0.6698,0.5757)
(0.6764,0.5687)
(0.6830,0.5615)
(0.6896,0.5542)
(0.6962,0.5468)
(0.7028,0.5392)
(0.7094,0.5316)
(0.7159,0.5239)
(0.7225,0.5160)
(0.7291,0.5081)
(0.7357,0.5000)
(0.7423,0.4918)
(0.7489,0.4836)
(0.7555,0.4752)
(0.7620,0.4667)
(0.7686,0.4581)
(0.7752,0.4494)
(0.7818,0.4405)
(0.7884,0.4316)
(0.7950,0.4226)
(0.8016,0.4134)
(0.8082,0.4042)
(0.8147,0.3948)
(0.8213,0.3853)
(0.8279,0.3757)
(0.8345,0.3661)
(0.8411,0.3563)
(0.8477,0.3464)
(0.8543,0.3363)
(0.8608,0.3262)
(0.8674,0.3160)
(0.8740,0.3056)
(0.8806,0.2952)
(0.8872,0.2846)
(0.8938,0.2739)
(0.9004,0.2632)
(0.9069,0.2523)
(0.9135,0.2413)

\rput[r](0.8200,0.8430){power law fit}
\PST@Dotted(0.8360,0.8430)
(0.9150,0.8430)

\PST@Dotted(0.2615,0.8663)
(0.2615,0.8663)
(0.2681,0.8603)
(0.2747,0.8543)
(0.2813,0.8484)
(0.2878,0.8424)
(0.2944,0.8365)
(0.3010,0.8305)
(0.3076,0.8246)
(0.3142,0.8186)
(0.3208,0.8126)
(0.3274,0.8067)
(0.3339,0.8007)
(0.3405,0.7948)
(0.3471,0.7888)
(0.3537,0.7829)
(0.3603,0.7769)
(0.3669,0.7710)
(0.3735,0.7650)
(0.3800,0.7590)
(0.3866,0.7531)
(0.3932,0.7471)
(0.3998,0.7412)
(0.4064,0.7352)
(0.4130,0.7293)
(0.4196,0.7233)
(0.4262,0.7173)
(0.4327,0.7114)
(0.4393,0.7054)
(0.4459,0.6995)
(0.4525,0.6935)
(0.4591,0.6876)
(0.4657,0.6816)
(0.4723,0.6756)
(0.4788,0.6697)
(0.4854,0.6637)
(0.4920,0.6578)
(0.4986,0.6518)
(0.5052,0.6459)
(0.5118,0.6399)
(0.5184,0.6340)
(0.5249,0.6280)
(0.5315,0.6220)
(0.5381,0.6161)
(0.5447,0.6101)
(0.5513,0.6042)
(0.5579,0.5982)
(0.5645,0.5923)
(0.5710,0.5863)
(0.5776,0.5803)
(0.5842,0.5744)
(0.5908,0.5684)
(0.5974,0.5625)
(0.6040,0.5565)
(0.6106,0.5506)
(0.6172,0.5446)
(0.6237,0.5386)
(0.6303,0.5327)
(0.6369,0.5267)
(0.6435,0.5208)
(0.6501,0.5148)
(0.6567,0.5089)
(0.6633,0.5029)
(0.6698,0.4969)
(0.6764,0.4910)
(0.6830,0.4850)
(0.6896,0.4791)
(0.6962,0.4731)
(0.7028,0.4672)
(0.7094,0.4612)
(0.7159,0.4553)
(0.7225,0.4493)
(0.7291,0.4433)
(0.7357,0.4374)
(0.7423,0.4314)
(0.7489,0.4255)
(0.7555,0.4195)
(0.7620,0.4136)
(0.7686,0.4076)
(0.7752,0.4016)
(0.7818,0.3957)
(0.7884,0.3897)
(0.7950,0.3838)
(0.8016,0.3778)
(0.8082,0.3719)
(0.8147,0.3659)
(0.8213,0.3599)
(0.8279,0.3540)
(0.8345,0.3480)
(0.8411,0.3421)
(0.8477,0.3361)
(0.8543,0.3302)
(0.8608,0.3242)
(0.8674,0.3183)
(0.8740,0.3123)
(0.8806,0.3063)
(0.8872,0.3004)
(0.8938,0.2944)
(0.9004,0.2885)
(0.9069,0.2825)
(0.9135,0.2766)

\PST@Border(0.1490,0.9680)
(0.1490,0.1260)
(0.9470,0.1260)
(0.9470,0.9680)
(0.1490,0.9680)

\catcode`@=12
\fi
\endpspicture
\end{center}
\caption[Distribution of national populations in the year
2000]{Distribution of national populations in the year 2005, plotted
  on a log-log scale. \citep{country-pop}. Also plotted are the best
  fits for a power law (slope -1.05) and lognormal ($\mu=14.8, \sigma=2.5$).}
\label{pop-hist}
\end{figure*}

We can rephrase the Chinese question in a different way: {\em What is
  the expected population size of one's country of birth}? It turns
out (see Fig \ref{pop-hist}) that there are far more countries with
fewer people, than countries with more people. The relationship
between population size and the number of countries looks roughly
proportional to $1/x$, where $x$ is the population of the country.
This law is an example of a power law, and it appears in all sorts of
circumstances, for example the frequency with which words are used in
the English language. 

With a $1/x$ power law, the number of countries of a given population size
exactly offsets the population of those countries, so anthropically
speaking, we should expect to find ourselves in just about any
sized country, with the same probability. Being in a country with a
population the size of Australia's would be no more
surprising than being in a more populated country such as the US or China.

However, the actual distribution of country populations turns out to
be a log normal distribution, \footnote{Thank you to Aaron Clauset for
pointing this out.}  whose probability distribution is
\begin{equation} p(x) = C/x \exp(-(\ln x -\mu)^2/2\sigma^2).
\end{equation} The parameters $\mu$ and $\sigma$ can be found by means
of the {\em maximum likelihood method} outlined by
\citet{Clauset-etal07}. In fact one can compare the likelihood of the
lognormal distribution explaining the population data with the
likelihood that the $1/x$ power law explains it, and it turns out to
be of the order of $10^{11}$ times as likely. Similar results hold for
other population datasets in the range 1965--2005. So indeed it would
be more likely for one to find oneself in a middle ranked country like
Kuwait or Estonia, than in the most populous nations of India or
China. However, the effect is not marked. India and China together
have about 2.4 billion people, and the total number of people living
in countries with populations in the range 10--100 million is about
2.1 billion, and in the range 100 million to a billion is about 1.5
billion.

Given the ubiquity of power laws, and the fact that a $1/x$ power law
exactly neuters any observer selection effect as in the above case,
might a $1/x$ power law be a signature of an arbitrary, or random
classification?

\section*{Mass distribution of animal species}

OK, well let's get back to our ants, and ask the question of what is
the expected abundance of our species, assuming we are randomly
sampled from all conscious species on the Earth. The distribution of species
populations tends to follow a power law, with a typical rank-abundance
plot within a species size class following a power law $A\propto r^m$ with exponent
$m=-1.9$ \citep{Siemann-etal99}, where $A$ is the abundance of the the $r$th
most abundant species. Rank-abundance plots are related to cumulative
size distributions \citep{Newman04}:
\begin{equation}\label{r(A)}
r(A) = N\int_A^\infty p(x) dx 
\end{equation}
where $N$ is the total number of species in a size class, and $p(x)$
the distribution of species abundances. Solving (\ref{r(A)}) implies
$p(A)$ is also a power law, with exponent $-1.52$. By the argument in
the previous section, we would therefore expect to find ourselves to be one
of the many species with few individuals, if all animals were
conscious, as the distribution of abundances falls off faster than
$1/A$. Yet our species abundance is many orders higher ($6\times10^9$)
than the minimum abundance for viability (approx $10^3$). However, for the most
part of our species' existence on the Earth, our abundance was much
less, and perhaps integrated over time, our total abundance is not so
different from that of other species of our size class.

However, let us ask a different question: ``what is our expected body
mass if we are randomly sampled from the reference class of conscious
beings?''. For this we need the abundance distribution $P(m)$ as a
function of body mass.

There is a well known biological law (called Damuth's
law) \citep{Damuth91}\index{Damuth's law} that states the population
density of a species is inversely proportional to the 3/4ths power of
that species' body mass, {\em i.e.} $A\propto m^{-3/4}$. To turn this
result into the mass distribution of individuals $P(m)$, we need to
multiply this law by the mass distribution of species $S(m)$.
Informally, we note that there are many more smaller bodied species of
animals than larger ones; there are many more types of insect than of
mammals, for example. The exact form of the distribution function
$S(m)$ is still a matter of conjecture. Some theoretical models suggest
that $S(m)$ is peaked at intermediate body
sizes \citep{Hutchinson-MacArthur59}, and experimental results appear to
confirm this \citep{Siemann-etal99}, although it must be admitted that
the latter study was confined to insects, and ignored the huge
diversity of nematodes. Of more interest was the finding that
$S(m)\propto P(m)^{0.5}$ \citep{Siemann-etal99} (Siemann {\em et al.}
use $I(m)$ instead of $P(m)$). Writing
\begin{equation}
P(m) \propto S(m)m^{-3/4} \propto P(m)^{1/2}m^{-3/4},
\end{equation}
we can solve for $P(m)$ as
\begin{equation}\label{mass distribution}
P(m)\propto m^{-3/2}.
\end{equation}
By the same arguments as above, we should expect to find ourselves
near the lower body mass of the class of conscious animals, ruling out
the vast majority of animals that are insects etc. 

\section*{Bayesian formulation}

The argument can be cast in a Bayesian framework in the following
way. Let $A$ represent the hypothesis that all animals are conscious,
and $B$ represent the observation that our observed body mass is
greater than (for arguments sake) 10kg. 

The previous argument could be criticised as suffering from what is
known as the ``Prosecutor's fallacy''. The value $p(B|A)$ can be
computed from (\ref{mass distribution}) by integration:
\begin{eqnarray}
p(B|A) &=& \int_{\mathrm{10kg}}^\infty P(m) dm \\
       &=& \left(\frac{\mathrm{10kg}}{m_0}\right)^{-1/2} \\ 
       &\approx& 10^{-5} \mathrm{\ with\ } m_0=1\mu\mathrm{g}  
\end{eqnarray}
where $m_0$ is the minimum mass of a conscious animal under hypothesis $A$. A
suitable choice for such an animal is {\em C. elegans}, a 1mm long
nematode with a nervous system consisting of 302 neurons. The mass of
an adult {C. elegans} is around 2$\mu$g \citep{Knight-etal02}. Even if
we were to limit the discussion to animals of the size of ants (our
titular species) or bigger (60$\mu$ g -- 2mg \citep{Kaspari05}),
$p(B|A) \approx 10^{-4}$.

However, the question we really want to know the answer to is what is
$p(A|B)$ --- what is the likelihood of all animals being conscious,
given that our observed body mass is more than 10kg?

Bayes law is written as
\begin{equation}\label{Bayes}
p(A|B) = \frac{p(B|A) p(A)}{p(B)}
\end{equation}
The term $p(A)$ represents our prior conviction in $A$. We can, for
the sake of argument, assume $p(A)=1$ here. Any lesser value only
increases the force of this argument.

The final term $p(B)$ is the probability of observing one's body mass
greater than 10kg. Since we don't know the mass distribution of
conscious beings, we cannot calculate this value directly. However, we
can use a form of anthropic reasoning introduced by \citet{Gott94}.
In that case, Gott argued that a number drawn at random from a uniform
distribution on the numbers $1\ldots N$ would find its value to lie in
the range $(0.5N,N]$ with confidence 95\%. Suppose instead that the
numbers were ordered according to some attribute $m$, drawn from some
unknown distribution $M(m)$. Then by the same argument, we can say
that with 95\% confidence
\begin{equation}\label{Mint}
\int_{m}^\infty M(m')dm' > 0.05.
\end{equation}
But the left hand side of (\ref{Mint}) with $m=10$kg is just our term
$p(B)$, where $M(m)$ is the unknown distribution of masses of
conscious observers. Thus with confidence $c\in[0,1)$, we can assume
$p(B)>1-c$. 

Plugging this into (\ref{Bayes}), our confidence in
hypothesis $A$ being wrong is
\begin{equation}\label{confidence}
c = 1-\sqrt{p(B|A)}
\end{equation}
which is 99.7\% for nematodes and 99\% for ants. By contrast, for the
proposition that all mammals are conscious, our confidence in this
being wrong by (\ref{confidence}) is only about 90\%, using the smallest
known mammal mass of about 2g for the Pygmy Shrew. 90\% is generally
considered not statistically significant, so anthropic reasoning
cannot be used to rule out the consciousness of all mammals without
further refinement of $p(A)$.

\section*{Conclusion}

In this paper, the {\em reference class} of anthropic reasoning
is used as a way to reason about the species of animals 
that could be conscious. Considering the reference class to be all
conscious animals on the Earth, one applies known distributions of
species abundances to determine that: one's nationality is not
expected to be any particular country, owing to a $1/x$ distribution
of population sizes; that one's body mass should be near the lower
limit of the set of conscious animals; and the abundance of one's own
species should be near the lower limit of species abundances. 

Considering our body mass is substantially higher than the average
animal (who is an insect, or even possibly a nematode), we can
conclude that the vast bulk of the animal kingdom is unlikely to be
conscious. We might also conclude, based on the high present abundance
of humans, that most species of our mass class are also not conscious,
since we should also expect to find ourselves near the lower limit of
species abundance of conscious species. But this would be a mistake
--- it is only natural, assuming we're born human, to be born in an
era of high human abundance. Integrated over our entire species
lifetime, the total human abundance may not be so different from that
of other species in our size class.


\end{document}